\documentstyle[a4wide,12pt,epsf]{article}
\begin{document}
\begin{flushright}
\baselineskip=12pt
{SUSX-TH-00-017}\\
{RHCPP 00-03T}\\
{hep-th/0011289}\\
{October  2000}
\end{flushright}
\def\IZ{Z\kern-.4em Z}
\begin{center}
{\LARGE \bf SUPERSYMMETRIC STANDARD MODELS ON D-BRANES \\}
\vglue 0.35cm
{D.BAILIN$^{\clubsuit}$ \footnote
{D.Bailin@sussex.ac.uk}, G. V. KRANIOTIS$^{\spadesuit}$ \footnote
 {G.Kraniotis@rhbnc.ac.uk} and A. LOVE$^{\spadesuit}$ \\}
	{$\clubsuit$ \it  Centre for Theoretical Physics, \\}
{\it University of Sussex,\\}
{\it Brighton BN1 9QJ, U.K. \\}
{$\spadesuit$ \it  Centre for Particle Physics , \\}
{\it Royal Holloway and Bedford New College, \\}
{\it  University of London,Egham, \\}
{\it Surrey TW20-0EX, U.K. \\}
\baselineskip=12pt

\vglue 0.25cm
ABSTRACT
\end{center}

{\rightskip=3pc
\leftskip=3pc
\noindent
\baselineskip=20pt
Type IIB superstring models with the standard model gauge group on 
D3-branes and with massless matter associated with open strings joining 
D3-branes to D3-branes or D3-branes to ${\rm D}7_3$-branes are studied.
Models with gauge coupling constant unification at an intermediate scale between about
$10^{10}$ and $10^{12}$GeV and consistency with the observed value of 
$\sin^2 \theta_W (M_Z)$ are obtained. Extra vector-like 
 states and extra pairs of Higgs doublets 
play a crucial role.}

\vfill\eject
\setcounter{page}{1}
\pagestyle{plain}
\baselineskip=14pt

Until recently, heterotic string theory has been the framework for 
model building in string theory. It was possible to construct three-generation 
models with realistic gauge groups in compactifications of the weakly 
coupled heterotic string on an orbifold. However, there was some difficulty 
in finding a natural explanation for the discrepancy between the 
``observed'' unification of gauge coupling constants at $2\times 10^{16}$ 
GeV \cite{DIMITRI} and the string scale of order $10^{18}$GeV where unification of coupling 
constants should occur at tree level \cite{mismatch}. These two distinct scales have been reconciled 
either by employing the moduli-dependent loop corrections to 
the gauge kinetic function with a value of the $T$-modulus 
an order of 
magnitude larger than that obtained from a straightforward minimization of the 
effective potential (in hidden sector gaugino 
condensate models) \cite{Tsize}, or by introducing extra 
matter states at an intermediate 
scale \cite{extramatter}, in addition to those of the minimal supersymmetric standard model 
(MSSM). In the strongly coupled heterotic string theory in the corner of 
$M$-theory corresponding to 11-dimensional supergravity at low energies, 
this problem is resolved differently through the existence of an extra 
dimension which allows the string scale to differ substantially from the 
 four-dimensional Planck scale \cite{witten}.

An alternative way forward, which has been the subject of recent 
interest, is to employ type IIB superstring theory compactified on 
an orientifold or orbifold with ${\rm D}$-branes 
present \cite{FERNANDO,INTER2}. In that case, the 
string scale can be 
altered by adjusting the ``radii'' of the 
underlying torus to obtain string scales which differ from the Planck 
scale. In this context, it is quite natural 
for the lack of supersymmetry in an anti-D-brane \cite{URAN} hidden sector 
to break supersymmetry in the observable sector by gravitational interactions between the two sectors.
Since we expect the 
lack of supersymmetry in the anti-D-brane sector to be characterized by 
the string scale $M_s$, the mass of 
sparticles in the observable sector will be of order $M_s^2/M_p$, 
where $M_p$ is the four-dimensional Planck mass. For sparticle masses 
of order $1$TeV we must have $M_s$ around $10^{10}$ to $10^{12}$GeV. Then, unification of 
gauge coupling constants at a scale of this order 
 \cite{INTER} is to be 
expected (apart from some subtleties to do with Kaluza-Klein modes and winding 
modes which we shall touch on later.) Interestingly, it is possible 
for such models to contain the extra matter required to allow the 
renormalization group equations to run to unification at this lower scale.

A recent approach to type IIB D-brane model building (the so called 
bottom-up approach) \cite{URANGA} has been to set up the observable sector gauge 
group and matter fields on a set of D-branes at a $R^6/\IZ_N$ orbifold 
singularity $before$ embedding this local theory in a global orbifold 
(or orientifold or Calabi-Yau) theory. The reason why this is an 
efficient approach to model building is that some properties of the 
model, such as the number of generations, depend only on the local 
theory. In this approach, it has been proved possible to obtain 
three-generation models consistent with the observed value of 
$\sin^2 \theta_W$ with D3-branes and D7-branes located at a  
$R^6/\IZ_3$ singularity and with the local theory embedded in a 
$\IZ_3$ orbifold, provided that the observable sector gauge group 
is the left-right symmetric $SU(3)\times SU(2)_L\times SU(2)_R \times U(1)$. 
In the examples studied with standard model gauge group 
$SU(3)\times SU(2) \times U(1)$ it did not prove possible to 
obtain consistency with the experimentally 
measured $\sin^2 \theta_W$. It is our purpose here to 
construct alternative $\IZ_3$ orbifold compactified type IIB ${\rm D}$-brane 
models which are consistent with the observed $\sin^2 \theta_W$ when the 
gauge group is that of the standard model. 

As in \cite{URANGA}, in the models we shall study here the gauge fields of the standard 
model are associated with a set of D3-branes at a $R^6/\IZ_3$ singularity, 
which we take to be the origin. The action of $\IZ_3$ on complex 
scalars is given by 
\begin{equation}
\theta={\rm diag}(e^{2\pi i b_i/3}, e^{2\pi i b_2/3},e^{2 \pi i b_3/3})
\end{equation}
with 
\begin{equation}
b_1=b_2=-1, b_3=2,
\end{equation}
which respects supersymmetry. The action of $\theta$ on the Chan-Paton 
indices of open strings ending on the D3-branes is given by
\begin{equation}
\gamma_{\theta,3}={\rm diag}(I_{n_0},\alpha I_{n_1}, \alpha^2 I_{n_2})
\end{equation}
where the $n_i$ are non-negative integers and 
\begin{equation}
\alpha=e^{2\pi i/3}
\end{equation}
$\IZ_3$ projections can then be made on the gauge field and matter 
field states. The projection on the gauge bosons gives the gauge group 
$U(n_0)\times U(n_1) \times U(n_2)$ and with the choice 
\begin{equation}
n_0=3, n_1=2, n_2=1
\label{values}
\end{equation}
we {\it have} the standard model gauge group (up to some $U(1)$ factors) 
$U(3)\times U(2) \times U(1)$. The projection on the massless matter 
states gives supermultiplets 
$3 [(n_0,\bar{n}_1)+(n_1,\bar{n}_2)+(n_2,\bar{n}_0)]$ which with the 
values $(\ref{values})$ of the $n_i$ is
\begin{equation}
3[(3,2)_{1/6}+(1,2)_{1/2}+(\bar{3},1)_{-2/3}]
\label{multi}
\end{equation}
where the weak hypercharge has been identified as 
\begin{equation}
Y=-\sum_{i=0}^{2}\frac{Q_{n_i}}{n_i}=-(\frac{1}{3}Q_{3}+\frac{1}{2}
Q_{2}+Q_{1})
\end{equation}

The model as it stands does not satisfy the twisted tadpole cancellation 
conditions for a $\IZ_3$ singularity 
\begin{equation}
3 {\rm Tr}\gamma_{\theta,3}-{\rm Tr}\gamma_{\theta,7_1}-
{\rm Tr}\gamma_{\theta,7_2}+{\rm Tr}\gamma_{\theta,7_3}=0
\label{tadpole}
\end{equation}
where the $\gamma_{\theta,7_i}$ are for the ${\rm D}7_i$-branes which 
overlap the origin, where the D3-branes are located. It is therefore 
necessary to introduce some D7-branes to achieve the cancellation. 
We choose to introduce only ${\rm D}7_3$-branes at complex coordinate 
$y_3=0$. The general action of $\theta$ on the Chan-Paton indices 
of the open strings ending on the ${\rm D}7_3$-branes is given by 
\begin{equation}
\gamma_{\theta,7_3}={\rm diag}(I_{u_0^3},\alpha I_{u_1^3},\alpha^2 I_{u_2^3})
\end{equation}
where the $u_i^3$ are non-negative integers. 
Then ($\ref{tadpole}$) becomes
\begin{equation}
(9+u_0^3)+(6+u_1^3)\alpha+(3+u_2^3)\alpha^2=0
\end{equation}
whose  general solution is 
\begin{equation}
u_0^3=u, u_1^3=3+u, u_2^3=6+u
\end{equation}
where $u$ is a non-negative integer, so that 
\begin{equation}
\gamma_{\theta,7_3}={\rm diag}(I_u,\alpha I_{3+u},\alpha^2 I_{6+u})
\end{equation}
The gauge group associated with the D$7_3$-branes is 
$U(u)\times U(3+u)\times U(6+u)$ and the associated massless states 
are in supermultiplets in representations of the gauge group
\begin{equation}
3[(u,\overline{3+u})+(3+u,{\overline{6+u}})+(6+u,\bar{u})]
\end{equation}

The  $3 7_3+7_3 3$ sector states, arising from open strings with one 
end on a ${\rm D}3$-brane and the other end on 
a ${\rm D}7_3$-brane, are in 
non-trivial representations of both the ${\rm D}3$-brane and 
the ${\rm D}7_3$-brane 
gauge groups and are 
\begin{eqnarray}
& &(3,\overline{3+u})_{-1/3}+(2,\overline{6+u})_{-1/2}+(1,\bar{u})_{-1} \nonumber \\
&+& (u,\bar{2})_{1/2}+(3+u,\bar{1})_1+(6+u,\bar{3})_{1/3}
\end{eqnarray}
 (If $u=0$ the representations $(1,\bar{u})_{-1}$ and $(u,\bar{2})_{1/2}$ are absent.) Combining
 the 33 and $37_3+7_3 3$ matter, the content in terms of representations of $SU(3)\times SU(2)\times U(1)$ is
\begin{eqnarray}
& & 3(Q_L+u_L^c+d_L^c+L+H_1+H_2+e_L^c) \nonumber \\
&+& (3+u)(d_L^c+\bar{d}_L^c)+u(H_1+H_2)+u(e_L^c+\bar{e}_L^c)
\label{yli2}
\end{eqnarray}
which is 3 generations plus some extra pairs of Higgs doublets 
and vector-like 
quark and lepton matter.

In order to reduce the size of the ${\rm D}7_3$-brane gauge group we specialize 
to 
\begin{equation}
u=3\tilde{u}
\end{equation}
where $\tilde{u}$ is a non-negative integer, and introduce a Wilson line 
in the second complex plane, embedded in the ${\rm D}7_3$-brane gauge group,
\begin{equation}
\gamma_{W,7_3}={\rm diag}(I_{\tilde{u}},\alpha I_{\tilde{u}}, 
\alpha^2 I_{\tilde{u}}, I_{1+\tilde{u}}, \alpha I_{1+\tilde{u}},
\alpha^2 I_{1+\tilde{u}},I_{2+\tilde{u}},\alpha I_{2+\tilde{u}}, 
\alpha^2 I_{2+\tilde{u}})
\end{equation}
This Wilson line may be represented by the shift
\begin{equation}
W_{7_3}=-\frac{1}{3}\Bigl(0^{\tilde{u}},1^{\tilde{u}},2^{\tilde{u}},
0^{1+\tilde{u}},1^{1+\tilde{u}},2^{1+\tilde{u}},0^{2+\tilde{u}},
1^{2+\tilde{u}},2^{2+\tilde{u}}\Bigr)
\end{equation}
There is then an additional projection on the root vectors $\rho_a$ for 
$7_3 7_3$ sector gauge fields and massless matter states
\begin{equation}
\rho_{a}. W_{7_3}=0 \;\;({\rm mod}\;\IZ)
\end{equation}
This breaks the ${\rm D}7_3$-brane gauge group to 
$[U(\tilde{u})\times U (1+\tilde{u})\times U (2+\tilde{u})]^3 $ and the 
$7_3 7_3$ sector matter states reduce to 
\begin{equation}
3 \Bigl[(\tilde{u},\overline{1+\tilde{u}})+(1+\tilde{u})(\overline{2+\tilde{u}})+
(2+\tilde{u},\bar{\tilde{u}})\Bigr]
\label{yli}
\end{equation}
for $each$ of the $U(\tilde{u})\times U (1+\tilde{u})\times U (2+\tilde{u})$ 
factors of the gauge group. The Wilson line does not delete any 
$3 7_3+7_3 3$ states because the D3-branes are at the origin and  
massless $3 7_3+7_3 3$ states have the ends of the string at the origin.
Thus, the influence of the Wilson line is not felt. Decomposing with 
respect to the $[U(\tilde{u})\times U (1+\tilde{u})\times U (2+\tilde{u})]^3 $ 
gauge group, the $3 7_3+7_3 3$ states are 
\begin{eqnarray}
& &(3, \overline{1+\tilde{u}})_{-1/3}+(2,\overline{2+\tilde{u}})_{-1/2}+
(1,\bar{\tilde{u}})_{-1} \nonumber \\
&+& (\tilde{u},\bar{2})_{1/2}+(1+\tilde{u},\bar{1})_1+
(2+\tilde{u},\bar{3})_{1/3}
\end{eqnarray}
for $each$ of the $U(\tilde{u})\times U (1+\tilde{u})\times U (2+\tilde{u})$ 
factors of the gauge group. Yukawa couplings of the 
$3 7_3+7_3 3$ states to $7_3 7_3$ states arise from the superpotential terms

\begin{eqnarray}
& & \phi^{7_3 7_3}_{(2+\tilde{u},\bar{\tilde{u}})}\phi^{3 7_3}_{
(2,\overline{2+\tilde{u}})}\phi^{7_3 3}_{(\tilde{u},\bar{2})} \nonumber \\
&+& \phi^{7_3 7_3}_{(\tilde{u},\overline{1+\tilde{u}})}
\phi^{3 7_3}_{(1,\bar{\tilde{u}})} \phi^{7_3 3}_{(1+\tilde{u},
\bar{1})}  \nonumber \\
&+& \phi^{7_3 7_3}_{(1+\tilde{u},\overline{2+\tilde{u}})}
\phi^{3 7_3}_{(3,\overline{1+\tilde{u}})}
\phi^{7_3 3}_{(2+\tilde{u},\bar{3})}
\label{superp}
\end{eqnarray}
for $each$ of the $U(\tilde{u})\times U (1+\tilde{u})\times U (2+\tilde{u})$ 
factors of the gauge group.
As can be seen from ($\ref{yli}$), there are three distinct 
$7_3 7_3$ sector multiplets for any given gauge group representations. Mass 
can be given to some (or all) of the extra vector-like matter states by 
turning on an expectation value for $7_3 7_3$ sector scalars. 
 We shall return to the implications for $\sin^2 \theta_W$ 
shortly. 

To complete the model we embed the $\IZ_3$ singularity at the origin 
in a  $\IZ_3$ orbifold. Because the D$7_3$-branes at $y_3=0$ not only overlap
the fixed point at the origin but also the fixed points at 
$(\pm 1,0,0),(0,\pm 1,0)$ and $(\pm 1,\pm 1,0)$, we must ensure that the 
twisted tadpole cancellation conditions are also satisfied at these 
fixed points. In the presence of the Wilson line in the second 
complex plane the twisted tadpole 
conditions are modified to 
\begin{equation}
3 {\rm Tr}\gamma_{\theta,3}+{\rm Tr}(\gamma_{\theta,7_3}
\gamma_{W,7_3}^{p})=0
\end{equation}
(assuming only D3-branes and D$7_3$-branes)  where $p=0$ for 
($\pm 1,0,0$) and $p=\pm 1$ for ($0,\pm 1,0$) and $(\pm 1, \pm 1,0)$. 
However,
\begin{equation}
{\rm Tr} (\gamma_{\theta,7_3} \gamma_{W,7_3}^{p})=0
\end{equation}
for $p=\pm 1$. Thus, the twisted tadpole conditions are already satisfied 
at $(0,\pm 1,0)$ and $(\pm 1,\pm 1,0)$. 
We can arrange for the  twisted tadpole conditions  to be satisfied 
at $(\pm 1,0,0)$ by including D3-branes at these fixed points. One suitable 
choice is to add 6 D3-branes with $\gamma_{\theta,3}$ of the same form 
\begin{equation}
\gamma_{\theta,3}={\rm diag}(I_3,\alpha I_2, \alpha^2 I_1)
\label{choice}
\end{equation}
as at the origin at each of the fixed points $(\pm 1,0,0)$. 
Another simple choice would be to add 3 D3-branes at each of the fixed points 
with 
\begin{equation}
\gamma_{\theta,3}={\rm diag}(I_2, \alpha I_1)
\end{equation}
For definiteness we considered the choice (\ref{choice}). 
There is then a total of 18 D3-branes and a total of 
$9+9 \tilde{u}$ D$7_3$-branes. Finally, to cancel untwisted tadpoles, 
it is necessary to add an equal number of $\overline{{\rm D}3}$-branes to 
balance the  D3-branes and an equal number of 
$\overline{{\rm D}7}_3$-branes to 
balance the ${\rm D}7_3$-branes. We would also like to avoid overlap of the 
anti-branes with the observable sector ${\rm D}3$-branes and 
${\rm D}7_3$-branes 
at the origin and overlapping the origin, respectively. This is easily 
achieved by placing the $\overline{{\rm D}3}$-branes and 
$\overline{{\rm D}7}_3$-branes at complex coordinate $y_3 \not =0$. For 
example, we might place 18  $\overline{{\rm D}3}$-branes with 
\begin{equation}
\gamma_{\theta,\bar{3}}={\rm diag}(I_6,\alpha I_6,\alpha^2 I_6)
\end{equation}
at a single fixed point with $y_{3}\not =0$, and $9+9\tilde{u}$ 
$\overline{{\rm D}7}_3$-branes  at $y_3=1$ or ($y_3=-1$) with 
\begin{equation}
\gamma_{\theta,\overline{7}_3}={\rm diag}(I_{3+3\tilde{u}}, 
\alpha I_{3+3\tilde{u}},\alpha^2 I_{3+3\tilde{u}})
\end{equation}
Because ${\rm Tr}\gamma_{\theta,\bar{3}}$ and ${\rm Tr}\gamma_{\theta,
\bar{7_3}}$ are both zero, the 
twisted tadpole conditions continue to be satisfied.
As a consequence of the geometrical separation of brane and 
anti-brane sectors, the anti-D-brane sectors, in which supersymmetry 
is absent, communicate only gravitationally with the observable 
D-brane sectors. As discussed earlier,  we then expect unification of gauge 
couplings at a scale of $10^{10}$ to $10^{12}$GeV.

We turn next to the consistency of such models with the observed value 
of $\sin^2 \theta_W$. In general, if we allow for running of the 
observable gauge coupling constants to a unification scale $M_X$, and if, 
for greater generality, we also allow that any ``light'' extra matter 
states, over and above those of the MSSM, have masses on a scale $M_Y$ 
between $M_Z$ and $M_X$, then, in a supersymmetric theory,
\begin{eqnarray}
\sin^2 \theta_W(M_Z)&=&\frac{3}{14}\Bigl[1+\frac{11}{6 \pi} \alpha(M_Z)
(b_2-\frac{3}{11}b_1)ln\frac{M_X}{M_Y} \nonumber \\
&+&\frac{11}{6\pi}
\alpha(M_Z)(\tilde{b}_2-\frac{3}{11}\tilde{b}_1)ln \frac{M_Y}{M_Z}\bigr]
\end{eqnarray}
and 
\begin{eqnarray}
\alpha^{-1}_3(M_Z)&=&\frac{3}{14}\Bigl[\alpha^{-1}(M_Z)-\frac{1}{2\pi}
(b_1+b_2-\frac{14}{3}b_3)ln\frac{M_X}{M_Y} \nonumber \\
&-&\frac{1}{2\pi}(\tilde{b}_1+\tilde{b}_2-\frac{14}{3}\tilde{b}_3)
ln\frac{M_Y}{M_Z}\Bigr]
\end{eqnarray}
or, equivalently,
\begin{eqnarray}
\sin^2 \theta_W(M_Z)&=&\frac{3}{14}+\frac{11}{3}
\frac{(b_2-\frac{3}{11}b_1)}{(b_1+b_2-\frac{14}{3}b_3)}\Bigl[
\frac{3}{14}-\alpha(M_Z)\alpha_3^{-1}(M_Z)\Bigr] \nonumber \\
&+&\frac{3}{14}\frac{11}{6\pi}\frac{\alpha(M_Z)}{
(b_1+b_2-\frac{14}{3}b_3)}\Bigl[(b_1+b_2-\frac{14}{3}b_3)(
\tilde{b_2}-\frac{3}{11}\tilde{b_1}) \nonumber \\
&-&(
b_2-\frac{3}{11}b_1)(\tilde{b}_1+\tilde{b}_2-\frac{14}{3}\tilde{b}_3)
\Bigr] ln (\frac{M_Y}{M_Z})
\end{eqnarray}
and
\begin{eqnarray}
\frac{3}{14}\frac{\alpha(M_Z)}{2\pi}ln \frac{M_X}{M_Y}&=&
(b_1+b_2-\frac{14}{3}b_3)^{-1}\Bigl[\frac{3}{14}
-\alpha(M_Z)\alpha_3^{-1}(M_Z)\Bigr] \nonumber \\
&-&\frac{3}{14}\frac{\alpha(M_Z)}{2\pi}\frac{(\tilde{b}_1+\tilde{b}_2-\frac{14}{3}\tilde{b}_3)}{(b_1+b_2-\frac{14}{3}b_3)}ln \frac{M_Y}{M_Z}
\end{eqnarray}
In these equations, $\frac{3}{14}$ is the value of the $\sin^2 
\theta_W$  at the unification scale in models 
where the D3-branes that carry the 
gauge group of the standard model are at a $\IZ_3$ singularity. 
The renormalization group coefficients for 
$SU(3)\times SU(2)\times U(1)$ are given in terms of Casimirs for the 
group factors $G_a$ and its massless matter field representations $R_a$ 
by
\begin{equation}
b_a=-3C_1(G_a)+\sum_{R_a}C_2(R_a), \;\;\;a=1,2,3
\end{equation}
The renormalization group coefficients for scales between $M_Z$ and 
$M_Y$, which we denote by $\tilde{b}_a$, are those of the MSSM.

So far as the observable sector is concerned, the massless matter 
content is as in (\ref{yli2}) with $u=3 \tilde{u}$. Let us first 
consider the case where we do not turn on any expectation values
 for the $7_3 7_3$ sector scalars, so that all of the extra matter in 
(\ref{yli2}), over and above that of MSSM, contributes to the running 
of the gauge coupling constants. If we do not introduce the extra 
scale $M_Y$ $(M_Y=M_Z)$, then
\begin{equation}
b_3=3\tilde{u},\;\;b_2=3(\tilde{u}+1),\;\;b_1=15+11\tilde{u}
\end{equation}
In that case, because $b_2-\frac{3}{11}b_1$ is negative, 
$\sin^2 \theta_W(M_Z)$ is less than $\frac{3}{14}=0.214$, whereas the 
observed value is 0.231. The correction due to the running of the 
gauge coupling constants is in the wrong direction. For the renormalization 
group coefficients of the MSSM, $\tilde{b}_2-\frac{3}{11}\tilde{b}_1$ is 
also negative so that including the extra scale $M_Y$ does not help.

Suppose next that the $7_3 7_3$ sector scalars give mass to $\alpha$ copies of $e_L^c+\bar{e}_L^c$, $\beta$ copies
of $H_1+H_2$, and $\gamma$ copies of $d_L^c+\bar{d}_L^c$ on a scale larger than $M_X$. 
This requires $\alpha \leq 3 \tilde {u}, \quad \beta < 3(\tilde{u}+1)$ and $\gamma \leq 3(\tilde{u}+1)$.
We use the values \cite{YELLOWREPORT} $\alpha^{-1} (M_Z)=128.9$ and $\alpha_3 (M_Z)=0.119$.
Then, if we do not
introduce an extra mass scale $M_Y$ (i.e. $M_Y=M_Z$), the best value for $\sin^2 \theta_W(M_Z)$ is $0.2275$, which occurs 
when $\alpha-\beta=2$ and $\gamma-\beta=4$. Moreover, the unification scale $M_X\simeq 1.3\times 10^{10}$GeV, which is
 consistent with observed supersymmetry breaking arising from gravitational interactions with an anti-brane sector. 
 To avoid $\alpha_3$ becoming infinite at a lower scale it is necessary 
to choose $\tilde{u}\leq 2$, 
but the above constraints on $\alpha, \beta$ and $\gamma$ can easily be satisfied.

Finally, let us suppose instead that expectation values of some $7_3 7_3$ sector 
scalars give mass to three copies of $d_L^c+\bar{d}_L^c$ and three copies 
of $e_L^c+\bar{e}_L^c$ on a scale larger than $M_X$.
Let us also suppose that some other $7_3 7_3$ sector scalars give mass to the extra matter,  over and above that 
of the MSSM,  on a scale $M_Y$ 
not very much larger than $M_Z$. 
Then
we obtain the observed value of
$\sin^2 \theta_W(M_Z)=0.231$ with 
\begin{equation}
\frac{M_Y}{M_Z}=13.9
\end{equation}
and 
\begin{equation}
M_X \sim 1.1 \times 10^{12}\; {\rm GeV}
\end{equation}

This  is consistent with supersymmetry breaking being transmitted 
gravitationally from an anti-brane sector provided the string scale is 
the unification scale. Assuming that the compactification is isotropic 
with a compactification scale $M_c$, then the string scale $M_s$ is 
given by \cite{aspects}
\begin{equation}
\frac{M_s^4}{M_c^3}=\alpha_{D3}\frac{M_p}{\sqrt{2}}
\end{equation}
where $\alpha_{{\rm D}3}$ is the value of $\frac{g^2}{4\pi}$ at unification 
when the observable gauge group is on ${\rm D}3$-branes and $M_p$ is the 
Planck mass. Winding modes have mass $M_w$ given by 
\begin{equation}
M_w=\frac{M_s^2}{M_c}
\end{equation}
To estimate the value of $\alpha_{{\rm D}3}$, we run the QCD fine structure 
constant $\alpha_3$ from $M_Z$ to $M_X \approx 1.1 \times 10^{12}$GeV. 
To avoid $\alpha_3$ becoming infinite at a lower scale it is necessary 
to choose $\tilde{u}\leq 1$. This restricts us to 
\begin{equation}
\tilde{u}=1
\end{equation}
because $\tilde{u}=0$ does not allow $7_3 7_3$ sector expectation 
values to give mass to any copies of $e_L^c+\bar{e}_L^c$. This can 
be seen from ($\ref{superp}$). 
With  the $\tilde{u}=1$ case, $\alpha_3$ runs only between $M_Z$ and $M_Y$ 
and its value at $M_X$ is approximately the same as at $M_Z$. 
Taking $M_s\approx 1.1 \times 10^{12}$GeV, then $M_w\sim 10^2 M_s$ and the 
scale associated with winding modes is above the string scale. The 
compactification scale $M_c$ is not directly relevant to unification 
because there are no Kaluza-Klein modes when all the boundary conditions 
for the compact manifold are Dirichlet. Thus, we may take the string scale 
to be the unification scale.

The situation with regard to mass hierarchies for the quark and lepton masses 
is the same as for the model discussed in \cite{URANGA}. Lepton mass terms require the 
coupling in the superpotential of the chiral fields $L,e^c_L$ and $H_1$ (possibly accompanied by
 some $7_3 7_3$ sector fields and/or $3^{'}7_3+7_3 3^{'}$ fields, which are uncharged with 
respect to the standard model gauge group; ($3^{'}$ refers to non-standard-model ${\rm D}3$-branes.)
Such couplings are allowed (as usual) by conservation of weak hypercharge, but they are $not$ allowed 
by conservation of the other two $U(1)$ charges originating from the standard model gauge group. It is 
expected that the global conservation of all $U(1)$ charges, including those originating from the 
${\rm D}7_3$-brane and ${\rm D}3^{'}$-brane gauge groups, will survive after the anomalous $U(1)$ gauge 
symmetries have been broken 
by a modified Green-Schwarz mechanism. For the lepton mass terms above it is obvious that 
$Q_2$ is not conserved, and the inclusion of other standard model gauge singlet fields does not 
alleviate the problem.  However, the global symmetries do allow the coupling 
\begin{equation}
Le^c_L \hat{H}_1=\phi^{3 7_3}_{(2,\overline{\tilde{u}+2})}\phi^{7_3 7_3}_{(\tilde{u}+2,\bar{\tilde{u}})}\phi^{7_37_3}_{(\tilde{u},\overline{\tilde{u}+1})}
\phi^{7_3 3}_{(\tilde{u}+1,\bar{1})}\phi^{3 7_3}_{(1,\bar{\tilde{u}})}\phi^{7_3 3}_{(\tilde{u},\bar{2})} M_s^{-3}
\label{lepmass}
\end{equation}
where $\hat{H}_1$ is an effective Higgs field 
\begin{equation}
\hat{H}_1=\phi^{7_3 7_3}_{(\tilde{u}+2,\bar{\tilde{u}})}\phi^{7_37_3}_{(\tilde{u},\overline{\tilde{u}+1})}
\phi^{3 7_3}_{(1,\bar{\tilde{u}})}\phi^{7_3 3}_{(\tilde{u},\bar{2})} M_s^{-3}
\end{equation}


A similar discussion shows that baryon number $B$ is (perturbatively) conserved in this model since $B=\frac{1}{3}Q_3$. Thus the 
proton is absolutely stable, and this is also the case in the model of reference \cite{URANGA}. On the other hand these global 
$U(1)$ symmetries do $not$ forbid the lepton-number non-conserving, dimension 4 operator $Q_L d^c_L L$, nor some lepton-number non-conserving, dimension 6 
operators. Some other resolution of the problem will have to be found. In this respect the present model cannot improve upon that of reference 
\cite{URANGA}.

It is interesting to note that the models discussed here possess
dual models in which the observable sector gauge group is on 
${\rm D}7_3$-branes at $y_3=0$ instead of D3-branes at the origin. 
So far as the observable 
sector matter is concerned ${\rm D}3$-branes and 
${\rm D}7_3$-branes are replaced by ${\rm D}7_3$-branes and ${\rm D}7_1$-branes.
(Strictly, we should recast the models we have been discussing in 
terms of D3-branes and D$7_2$-branes to obtain the duality.) The 
dual models have action on the Chan-Paton indices
\begin{equation}
\gamma_{\theta,7_3}={\rm diag}(I_3,\alpha I_2,\alpha^2 I_1)
\end{equation}
for ${\rm D}7_3$-branes at $y_3=0$ and 
\begin{equation}
\gamma_{\theta,7_1}=-{\rm diag}(\alpha^2 I_{2+\tilde{u}},I_{\tilde{u}}, 
\alpha I_{\tilde{u}+1})
\end{equation}
for ${\rm D}7_1$-branes at $y_1=0,\pm 1$. 
Notice that $9+9\tilde{u}$ ${\rm D}7_3$-branes all at $y_3=0$ with the 
addition of a Wilson line, in the original version, is replaced by 
$3+3\tilde{u}$ ${\rm D}7_1$-branes at each of $y_1=0,\pm 1$ with no 
Wilson line, in the dual version.
As before, the embedding in an orbifold is more model dependent.

In conclusion, we have constructed type IIB D-brane models with the 
standard model gauge group on D3-branes and the massless matter states 
for the standard model in 33 and $3 7_3+7_3 3$ sectors. 
These  models allow unification of gauge coupling constants at an 
intermediate scale of order $10^{10}$ to $10^{12}$GeV, consistently with the observed 
value of $\sin^2 \theta_W(M_Z)$. Additional vector-like states and 
pairs of Higgs doublets with low mass 
compared to the unification scale play 
a crucial role. 

\section*{Acknowledgements}
This research is supported in part by PPARC. We are grateful to Gerardo Aldazabal, Luis Ib$\rm {\acute{a}}\rm{\tilde{n}}$ez, 
 Fernando Quevedo and Angel Uranga for helpful discussions.

\end{document}